\documentclass{ws-procs975x65}
 
\def\eqnb{\begin{equation}}
\def\PTP{Prog. Theor. Phys.(Kyoto)}

\def\NPB{{Nucl. Phys.} {\bf B}}
\def\PLB{{Phys. Lett.} B}

\def\PRD{{Phys. Rev.} D}

\begin{document}

\title{CHIRAL SYMMETRY AND BRST SYMMETRY BREAKING, QUATERNION REALITY AND THE LATTICE SIMULATION
}

\author{Sadataka FURUI}

\address{School of Science and Engineering, Teikyo University,\\
Utsunomiya, 320-8551, Japan\\
E-mail: furui@umb.teikyo-u.ac.jp}

\begin{abstract}
We discuss that the deviation of the Kugo-Ojima color confinement parameter $u(0)$ from -1 in the case of quenched lattice simulation and the consistency with -1 in the case of full QCD simulation could be attributed to the boundary condition defined by fermions inside the region of $r<1$fm.
By using the domain wall fermion propagator in lattice simulation, we show that the chiral symmetry breaking in the infrared can become manifest when one assumes that the left-handed fermion on the left wall and  the right-handed fermion
on the right wall are correlated by a self-dual gauge field.
The relation between the infrared fixed point of the running coupling measured in lattice simulations, the prediction of the BLM renormalization theory,
the conformal field theory with use of the t'Hooft anomaly matching condition in non-SUSY supersymmetric theory and the quaternion real condition are discussed.
\end{abstract}


\bodymatter

\section{Introduction}
The infrared (IR) QCD is characterized by the color confinement and the chiral symmetry breaking. 
We performed lattice simulations\cite{SF09,sf08b} using the gauge configuration produced with domain wall fermion \cite{AABB07} which preserves the chiral symmetry in the zero - mass limit and compared with results of staggered fermion.  
In these lattice simulations and in comparison with other works, we observed qualitative differences between quenched and unquenched simulations.

\subsection{BRST symmetry}
The Kugo-Ojima color confinement attracted renewed interest by a conjecture of
 possible BRST (Becchi-Rouet-Stora-Tyutin) symmetry breaking in the infrared \cite{Dudal09,Kondo09}. These authors pay attention to the restriction of the gauge configuration to the fundamental modular region defined by Zwanziger to solve the Gribov problem. 

Kondo\cite{Kondo09} parametrized the horizon function
\begin{equation}
\langle (g f^{abc}A^b_\mu c^c)(g f^{def}A^e_\nu\bar c^f)\rangle_k=-\delta^{ad}\left(\delta^T_{\mu\nu}u(q^2)+F(q^2)\frac{q_\mu q_\nu}{q^2}(u(q^2)+w(q^2))\right)
\end{equation}\label{KO}
\noindent and assuming $w(0)=0$, obtained $u(0)=-2/3$.
There is an argument against this result \cite{Orsay09} showing that $w$ and $u$ are not multicatively renormalizable.  On the lattice, however the first term of the r.h.s of eq.(\ref{KO}) i.e. $\delta^{ab}\delta^T_{\mu\nu}u(q^2)$ is $\delta^{ab}\delta^T_{\mu\nu}S(U_\mu)u(q^2)$ where $S(U_\mu)$ is 
$
S(e^{A_\mu})=\frac{A_\mu/2}{{\rm tanh}{A_\mu/2}}
$
in the log$-U$ definition and
$
S(U_\mu)_{ab}=tr\left(\lambda^\dagger_a \frac{1}{2}\left\{\frac{U_\mu+U_\mu^\dagger}{2},\lambda_b\right\}|_{traceless\,p}\right)
$
in the $U-$linear definition.

The expectation value of $S$ is proportional to $e$
and $e/d$ is about 0.95 in $56^4$ lattice quenched SU(3) log$-U$ definition but
about 0.88 in the $U-$linear definition \cite{NF05}. 
When $S(0)\sim 0.85$ instead of 1 as in DSE, a solution of $w(0)= 0$ and 
$
u(0)=-\frac{2}{3}
$
is possible. 

\subsection{QCD running coupling}
Recently Appelquist\cite{App08} claimed by performing lattice simulation of running coupling in the Schr\"odinger functional method  that there is a critical number of flavors $N_f^c$ below which both chiral symmetry breaking and confinement set in. He assigned $8\leq N_f^c\leq 12$ and in the case of $N_f=8$ the running coupling monotonically increase as $\beta$ decreases and in the case of $16^4$ lattice and $\beta\sim 4.65$, ${\bar g^2(L)}/{4\pi}\sim 1.7$, and that there is no sign of infrared fixed point at $N_f<8$. However, applicability of the Schr\"odinger functional method in the region of $q$ below and around $\Lambda\sim 213\pm 40$MeV is not clear. Brodsky et al, \cite{BS08} argue that in the DSE  $q<\Lambda$ or $r>1$fm, confinement is essential and the quark-gluon should not be treated as free. 

In our lattice simulation of $N_f=2+1$ DWF, $\alpha_s(q)\sim 1.7$ at $q\sim 0.6$GeV and the deviation of the running coupling from 2-loop perturbative calculation is significant below $q\sim 3$GeV due to $A^2$ condensates. 
The running coupling data of JLab\cite{DBCK08} suggest presence of the infrared fixed point and that $N_f=3$ is close to the $N_f^c$. 

\begin{figure}
\begin{minipage}[b]{0.47\linewidth}
\begin{center}
\includegraphics[width=6cm,angle=0,clip]{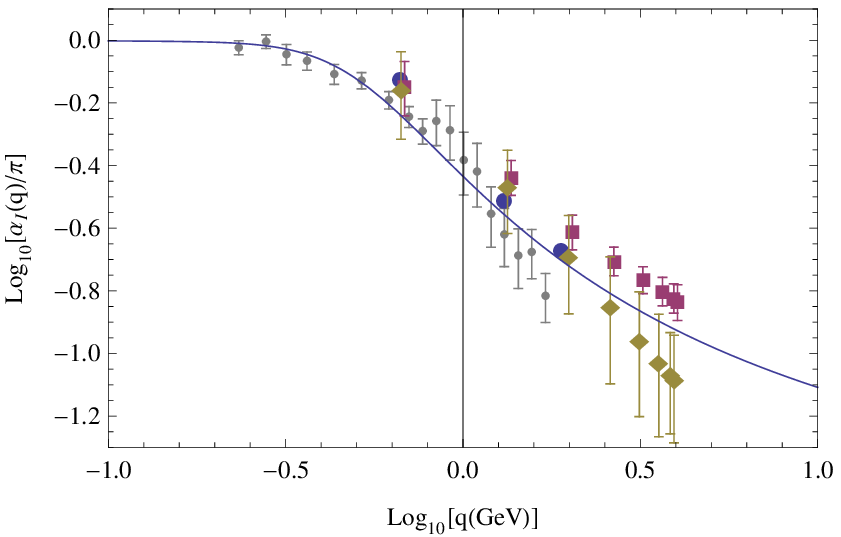}
\caption{The running coupling of the domain wall fermion. Coulomb gauge gluon-ghost coupling of $m_u=0.01/a{\rm (square)}$, $0.02/a{\rm (diamond)}$, and quark-gluon coupling of $m_u=0.01/a{\rm (large\, disks)}$. Small disks are the $\alpha_{s,g_1}$ derived from the spin structure function of the JLab group\cite{DBCK08}and the solid curve is their fit.}\label{alp_plt}
\end{center}
\end{minipage}
\hfill
\begin{minipage}[b]{0.47\linewidth}
\begin{center}
\includegraphics[width=6cm,angle=0,clip]{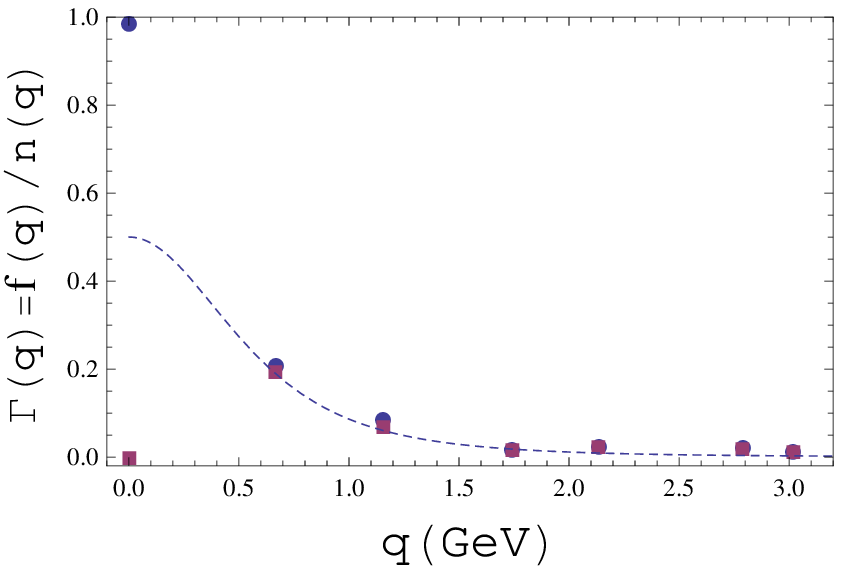}
\caption{The form factor of a proton using the DWF $m_u=0.01/a$ gauge fixed such that the fermion on the left wall and the right wall are correlated by self-dual gauge field, and the dipole fit with $M^2=0.71$GeV$^2$. At zero momentum left-handed(LH) fermion dominates the form factor, and at finite momentum, LH and RH contribute almost equally.}\label{charge_plt}
\end{center}
\end{minipage}
\end{figure}

\subsection{Quaternion real condition and proton form factor}
The infrared fluctuation of the gluon propagator could be attributed to the chiral anaomly. t'Hooft discussed that the anomaly cancellation on the color triplet and color sextet components in the 3 quark baryon sector\cite{tH80}, and formulations in non-SUSY supersymmetric theory were discussed by Sannino \cite{San09}. 
In standard DWF analysis, the limit of the distance between the left domain wall and the right domain wall approaching infinity corresponds to the continuum limit. I take a specific gauge that the fermion on the left domain wall and the right domain wall are correlated by a self-dual gauge field (instanton) that satisfy quaternion real condition\cite{CG81}. Then I calculate the ratio of the three point function and the two point function of fermions using the SU(6) proton wave function. As shown in Fig.2, at zero momenrtum the form factor comes exclusively from the left-handed fermion. 

A product of quaternions makes an octonion, whose automorphism is the exceptional lie group $G_2$ which has 3+6+5 dimensional stable manifolds and posesses the triality symmetry. The spontaneous supersymmetry breaking of massless fermions on 5 dim manifolds and ghost, gluon on 3 dim and 6 dim color space, and  the problem of the large critical $N_f$ for the conformality by about a factor of 3, would be explained by the triality symmetry of the $G_2$ group of the octonion\cite{Cartan66}. 
 Whether the large critical $N_f$ is an artefact, and how sensitive the results on the boundary conditions are further to be investigated. 

\section{Acknowledgments}
I thank Dr. F. Sannino, Prof. S. Brodsky for helpful information and discussion , and Dr. A. Deur for sending me the JLab experimental data. Numerical calculation was done at KEK, YITP Kyoto Univ. and  RCNP Osaka Univ.

\bibliographystyle{ws-procs975x65}
\bibliography{ws-pro-sample}

\end{document}